# An Interface using SOA Framework
# For Mediclaim Provider

**S. Nirmala Sugirtha Rajini**
**Dr.M.G.R. Educational and Research Institute, Chennai, India**
**Dr. T. Bhuvaneswari**
**Dr.M.G.R. Educational and Research Institute, Chennai, India**

**ABSTRACT:** SOA brought new opportunities for the long expected agility, reuse and the adaptive capability of information technology to the ever changing business requirements and environments. The purpose of this paper is to describe the implementation of Medical Insurance Claim Process Model using SOA. We adopt Service Oriented Architecture (SOA) to reduce the complexity among systems and solve data consistency problems among services. We choose n-tier and Service-Oriented Architecture (SOA) as our system environment. This model can also establish a potentially new innovative market branch for the insurance industry.

## Introduction

SOA is a design pattern composing of loosely coupled, discoverable, reusable, inter-operable platform independent services in which each of these services follow a well defined standard. Each of these services can be bound or unbound at any time and as needed. The foundation of service-oriented architecture (SOA) and the virtualization of services that are made available over the internet, gradually become an effective and scalable service delivery and consumption platform [LZ10].

Service-Oriented Architecture is a software architectural style that is employed for realizing and constructing business processes, which are composed of services [YCC08]. Service-oriented technology could extend

Information and Communication Technologies (ICT) to provide various services, which sometimes require a large amount of data exchange. The services can be deployed over the Internet, and can be combined to be re-used for new applications. Through the SOA platform, services could be delivered to end-users.

SOA enables reusability of software components, provides protocol independence, and facilitates application integration. It enforces basic software architecture principles such as modular design, abstraction, and encapsulation. Web services are based on open standards—in particular, XML and SOAP. These standards aim to achieve interoperability between applications implemented in different languages, running on diverse computer systems, and communicating over a network. It supports diverse processing efficiently and effectively, enabling cross-platform communication, and dynamically adapts to meet the changing needs.

## 1. Proposed Layered Architecture for Medical Insurance Claim Process Model using SOA

Web Services offer a new distribution channel for the insurance company, with more flexibility in offering policies and services from the available products. Web Services make it relatively easy for a user to use different medical policies. Web Service will make it easier for various types of users to offer products from an insurance company so that more users will be attracted towards insurance market.

The new possibilities offered by Web Services serve to render a lot of information to the user. The landscape of the insurance business will change with a wider use of Web Services, so that all web services involved will be able to communicate with each other. Although all communication is based on the same protocols, still there will be a need for an overall XML Schema that strictly determines the format of this communication. This XML Schema ensures that every web services involved in the same process 'talks the same language. Without using SOA concept in the Medical Claim process Model, few web services are needed to repeat for a number of times depending on the kinds of policies.

The following figure shows the medical claim process without using SOA concept.

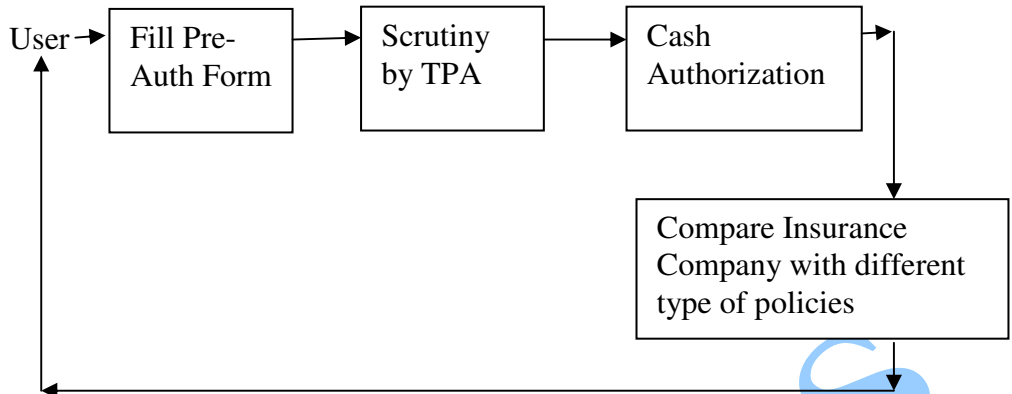

**Figure 1. Medical claim process without using SOA Concept**

Using SOA concept in the Medical Claim process Model, the web services like fill Pre-Authorization form, Scrutiny by TPA and cash Authorization services are needed only once instead of multiple times.

The following figure shows the layered architecture for medical insurance claim process Model with SOA concept. This architecture contains five layers.

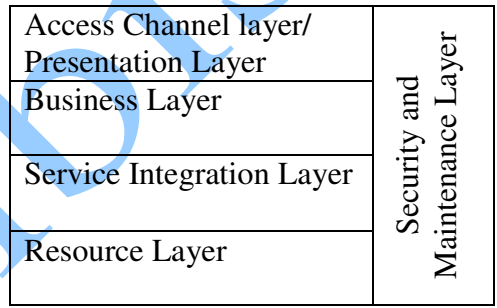

**Figure 2. Layered Architecture for Medical claim Insurance Process model with service integration layer**

**a) Access Channel Layer**

The Access Channel layer is responsible for handling all the interactions of various users with the web services and give the users a rich internet application experience through its user interface. This layer handles the construction and presentation of the user views, device specific handling of content and presentation, manages view states and transitions, controls

**11**

the user access and actions, manages the session and provides the right content for the right person through personalized content and customized look and feel [SRS10].

**b) Business Layer**

The Business layer comprises of application components, application services and application business processes that realize an application specific business function or process [SRS10].

**c) Service Integration Layer**

The Service Integration layer is to abstract the business layer (business process composition, services and components layer) from the underlying data sources and backend systems thereby providing transparent integration [SRS10]. This abstraction and transparency is enabled through the use of various adapters that span across various technologies, applications and custom-built.

**d) Resource Layer**

The resource layer is the layer between the business layer and a database or external services. The data access logic is responsible for persisting business entities to a database and retrieving individual business entities or sets of business entities on behalf of the business layer. The resource access layer may also contain service agents which are responsible for contacting other services to retrieve resources. The resource access layer should encapsulate all the code that deals with external resources, without leaking any of these implementation details to higher layers.

**e) Security and Quality Maintenance Layer**

This is mainly a background process. This layer mainly provides the capabilities for monitoring and maintaining quality for security, performance and availability. It uses sense and response mechanisms, and monitors SOA during their runtime and the important standards implementation of web service management.

## 2. A case study: Medical Insurance Claim Process Model

Insurance is a financial service allowing individuals or organizations "to pool their exposure to risks of economic loss resulting from the occurrence of uncontrollable events such as fire, death, earthquakes etc." [LZ10] Essentially, an individual or an organization pays a premium to the insurer that buys the guarantee of compensation for the losses due to the occurrence of those predefined set of risky events in the insurance policy.

MediClaim Insurance contains various Health policies which benefits us in different situations. Medical Insurance Claim Process Model help the insurance company to manage potentially health related events as a form of degree of injuries.

For the new user interface, we adopt web-based user environment. The web user interface server (WebUI) is the server which generates web page to the user browser. The WebUI handles the user browser requests directly, and then processes authentication calls, central SOAP calls, and data exchange calls if required. A user can use the Web server user interface to access the insurance module from a desktop computer or a PDA

We can avail the cash facility on hospitalization in case we can go for emergency at any of the network hospitals. The list of the network hospitals will be provided to the user when the user receives their health card from third party administrator (TPA). The insured will receive the Claims process is simplified.

On the event of hospitalization at any of the TPA's network hospitals, we don't have to pay for the treatment. The payment for emergency / or Surgical Benefits would be done by the TPA directly to the network hospital. These benefits paid will be as per your Insurance Policy provisions and the actual hospital expenditure incurred.

In our example we can implement the SOA concept in medical insurance claim process. It contains number of web services like authorization form service, scrutiny by TPA service, cash authorization service and cash payment service. Web Services is one of the most active and widely adopted implementation of SOA. It is based on an interoperable protocol called SOAP (Simple Object Access Protocol) and all communication between the server to client, client to client or server to server and in general application to application, uses the same protocol.

## 3. Process flow for Medical Insurance Claim Process Model

In a planned Hospitalization, the policyholder is required to fill the request for pre-authorization form along with the details of the illness and proposed treatment and expenses, certified by the treating doctor of the hospital, along with the user identification number and send through internet from the hospital.

The user identification number compares with the different type of insurance company policy databases. If there is no match with the company databases the user gets the message "identification number is invalid".

Otherwise they use the web services and directly issue the amount to the hospital. In case if the actual expense incurred is less than the amount eligible in policy provisions, the user can receive the difference from the TPA(Third Party Administrator).

The following figure shows the Medical Insurance Claim Process Model using SOA concept.

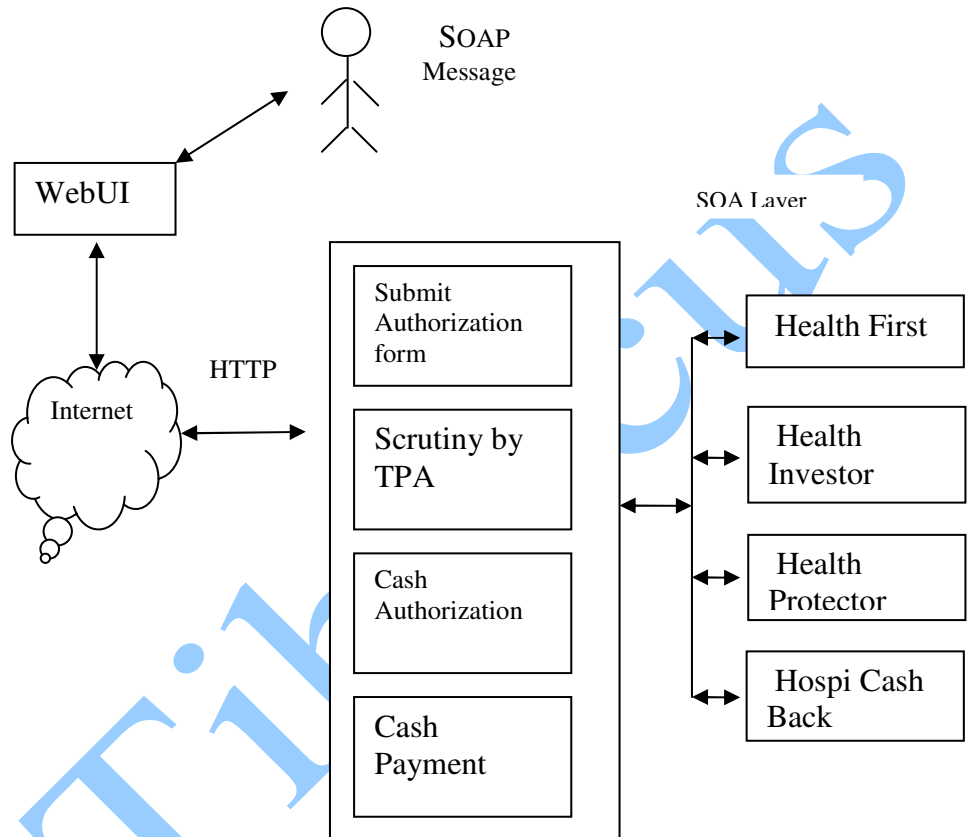

**Figure 3. Medical Insurance Claim Process Model Using SOA Concept**

## 4. Features of proposed system

The solution provides an efficient medical claim process, easy to use any time access from multiple locations and from heterogeneous technology environments. Performance, reliability, security requirements are addressed by the solution. As service-oriented architecture is quickly becoming a dominant paradigm for enterprise to effectively utilize IT investments, reduce the infrastructure and maintenance cost, their core business competencies and growth, third party business service that are openly available and are more focused.

**Conclusion**

The successful implementation of this model could encourage service consumers' confidence in utilizing the many available services from the vast service oriented environment, therefore truly able to enjoy the much expected benefits. On the implementation of SOA, reusability, maintainability of services, and easy integration, interoperability with diverse platforms on which the policy providers and users application systems are generated more. In future we can apply this concept to implement for the different policy providers with different issues.

**References**

[KMM08] **Firat Kart, Louise E. Moser, P. Michael Melliar-Smith** - *E-Healthcare System Using SOA,* 2008.

[LZ10] **Min Luo, Liang-Jie Zhang** - *An Insurance Model for Guaranteeing Service Assurance*, Integrity and QoS in Cloud Computing, 2010 IEEE International Conference on Web Services.

[SM09] **Asadullah Shaikh, Muniba Memon -** *The Role of Service Oriented Architecture* in Telemedicine Healthcare System 2009, International Conference on Complex, Intelligent and Software Intensive Systems.

[SNK09] **Stefan Seedorf, Khrystyna Nordheimer, Simone Krug -** *STraS: A Framework for Semantic Traceability* in Enterprise-wide SOA Life-cycle Management, 2009.

[SRS10] **G. Subrahmanya, V. R. K. Rao, Karthik Sundararaman -** Dhatri – *A Pervasive Cloud Initiative for Primary Healthcare Services*, 2010.

[VM06]   **Eugen Vasilescu, Seong K. Mun -** *Service Oriented Architecture (SOA) for Large Scale Distributed Health Care Enterprises*, Proceedings of the 1st Distributed Diagnosis and Home Healthcare (D2H2) Conference, Arlington, Virginia, USA, April 2-4, 2006.

[WWW]    **www.tata-aig-life.com**

[YCC08]  **Chi-Lu Yang, Yeim-Kuan Chang, Chih-Ping Chu -** *A Gateway Design for Message Passing on The SOA Healthcare Platform,* 2008 IEEE International Symposium on Service-Oriented System Engineering.